# Pressure dependence of the Boson peak in glassy $As_2S_3$ studied by Raman Scattering

K. S. Andrikopoulos[1,2,*], D. Christofilos[1], G. A. Kourouklis[1], S. N. Yannopoulos[2,*]

[1] *Physics Division, School of Technology, Aristotle University of Thessaloniki, GR-54124, Thessaloniki, Greece*

[2] *Foundation for Research and Technology Hellas Institute of Chemical Engineering and High Temperature Chemical Processes (FORTH / ICE-HT), P.O. Box 1414, GR-26504, Patras, Greece*

**Abstract**

A detailed pressure-dependence study of the low-energy excitations of glassy $As_2S_3$ is reported over a wide pressure range, up to 10 GPa. The spectral features of Boson peak are analysed as a function of pressure. Pressure effects on the Boson peak are manifested as an appreciable shift of its frequency to higher values, a suppression of its intensity, as well as a noticeable change of its asymmetry leading to a more symmetric shape at high pressures. The pressure-induced Boson peak frequency shift agrees very well with the predictions of the soft potential model over the whole pressure range studied. As regards the pressure dependence of the Boson peak intensity, the situation is more complicated. It is proposed that in order to reach proper conclusions the corresponding dependence of the Debye density of states must also be considered. Employing a comparison of the low energy modes of the crystalline counterpart of $As_2S_3$ as well as the experimental data concerning the pressure dependencies of the Boson peak frequency and intensity, structural or glass-to-glass transition seems to occur at the pressure ~4 GPa related to a change of local structure. Finally, the pressure-induced shape changes of the Boson peak can be traced back to the very details of the excess (over the Debye contribution) vibrational density of states.

1.  **Introduction**

The low energy spectrum of inelastically scattered light or neutrons in non-crystalline phases is dominated by scattering mechanisms absent form the parent crystal phase. Actually, the same spectral region in crystals is usually dominated by the Debye contribution; the corresponding vibrational density of states (VDoS), $g^D(\omega)$, is proportional to $\omega^2$ and is determined by the sound waves. The excess of the VDoS in a glass with respect to that of the crystal is considered as one of the universal characteristics of glasses [1], which together with other low temperature anomalies is also found in a number of disordered systems [2]. Numerous studies of glasses have shown that the nature of excitations that contribute to the VDoS, $g(\omega)$, in the low frequency region of the spectrum may originate either from low lying optical or transverse acoustic modes of parental crystals, see for example Ref. [3] for glasses with two-dimensional (2D) local structure and Ref. [4] for 3D network glasses, or from purely disorder-induced localised vibrational modes e.g. in metallic glasses [5]. Of course these are the two extreme cases and in many circumstances the origin of the Boson peak could be a combination of both or of a different mechanism. In essence, all modes that appear in the low energy region and contribute to a small, usually hardly observed bump in the VDoS, superimposed to the Debye level, would give rise to a strong peak in the experimental Raman and inelastic neutron scattering data. Thus, in the Raman spectra this excess in VDoS appears as a strong asymmetric peak slowly decaying towards higher wavenumbers. The term Boson peak (BP) was assigned to this peak owing to the fact that its intensity follows the predictions of the Bose law for the thermal population of the vibrational energy levels.

In order to understand the nature and origin of the Boson peak at the microscopic level, one has to study its spectral dependence on the glass composition or in the presence of external stimuli. Temperature is such a stimulus that can be rather easily applied and hence it has been employed as a control parameter in the vast majority of experimental and simulation studies. On the contrary, although pressure may induce significant changes in the vibrational excitations in glasses its use has been very limited due to inherent experimental difficulties in resolving the BP. However, the high range of pressures achieved in Raman experiments is of high importance for the verification of theories that provide exact prediction about the pressure dependence of the spectral features of the BP.

It is interesting to note that in all glasses experimentally studied up to now (see for example Refs. [6-9] for some representative publications) the general conclusion is that the application of pressure induces a shift of the BP position to higher energies and at the same time a noticeable suppression of its intensity. On the theoretical/simulation side, the influence of pressure or densification on BP has also been considered [10-13] showing similar effects to the experiment findings. One of the most powerful theoretical approaches dealing with the pressure dependence of the BP is the soft potential model [10], which predicts in detail the pressure dependence of the BP energy and intensity. Comparison of those predictions with experimental data shows a rather good agreement. However, in most studies and especially in polymeric glasses the pressure range is very limited and hence the predictions of the model [10] cannot be fully appreciated. In this context, pressure dependence studies of inorganic glasses are highly desirable in order to clarify the validity of the results of theoretical and simulations approaches.

In view of the above, we have undertaken a detailed pressure dependence Raman scattering study of the low energy excitations of glassy $As_2S_3$ whose locally 2D structure is highly amenable to pressure-induced changes. Previous studies of pressure effects on g-$As_2S_3$ include a study of the high frequency modes [14, 15] and a preliminary study of the BP restricted to low pressures [16]. In the present paper, we pay particular attention to the comparison of the results obtained in this work with recent theoretical approaches.

## 2. Experimental

For the preparation of glassy *$As_2S_3$*, elemental high purity (5N), multiply sublimed, *S* and *As* in pre-weighted quantities were placed in a quartz tube. The tube was evacuated, flame-sealed and placed for 48 hours in an oven at 750 $^o$C where it was frequently shaken to ensure homogenization. Finally, it was quenched to ambient temperature and annealed just below the glass transition temperature for several hours. The produced glass was homogeneous, transparent and bubble free.

Raman spectra were recorded with the aid of a micro-Raman instrumentation (triple monochromator working in the double subtractive configuration) and a cryogenic CCD detector. Excitation was achieved by a Ti-Sapphire laser (pumped by an $Ar^+$ laser) tuned at 780 nm. This wavelength ensured sub-bandgap excitation up to a pressure of several GPa [17] since the bandgap diminishes with increasing pressure. Thus, photoinduced phenomena associated with the illumination of amorphous $As_2S_3$ were abstained. The spectra were recorded in the backscattering geometry using a microscope (20x ULWD objective).

High pressure was applied using a diamond anvil cell of Mao and Bell type. The ~150 μm diameter of the gasket hole was filled with a minute piece of *$As_2S_3$*, chips of ruby (that enabled the calculation of pressure) and the 4:1 methanol-ethanol mixture (that served as the hydrostatic pressure-transmitting medium, ensuring hydrostatic conditions up to ~11 GPa). After each step of pressure application the sample was allowed to relax to the new pressure for several hours before the collection of the Raman spectrum. Spectra of good quality were easily recorded up to ~ 6GPa. From that point the signal-to-noise ratio gradually deteriorated as pressure was increased up to ~10 GPa, which was the maximum pressure applied. During the procedure of slow pressure release spectra were recorded following the aforementioned conditions in order to check for possible hysteresis phenomena.

3.  **Results**

Full scale, Stokes-side Raman spectra of g-$As_2S_3$ are presented in Fig. 1 for representative pressures. The spectra were scaled in relation to the integrated intensity of the spectral range [270-420] $cm^{-1}$. This envelope contains intramolecular vibrational bands of the $AsS_3$ pyramidal units that constitute the structure of g-$As_2S_3$ [18]. The scaling has been performed under the approximation that pressure application (at least up to the upper limit of the current work) does not induce appreciable modifications of intramolecular (covalent) bonds.

At least two additional bands exist in the spectral region of interest (Fig. 1), at ~188 $cm^{-1}$ and ~234 $cm^{-1}$. These bands are characteristic of molecular-like $As_4S_4$ species that manifest chemically broken order and inevitably exist at small concentration in this stoichiometric glass [19]. The existence of these species justifies the presence of the S-S vibrational band at ~490 $cm^{-1}$, also present in the glass spectrum but not in the spectrum of the crystal. Figure 1 reveals that the spectral features associated with the BP appear to be significantly influenced by the application of pressure. At first, there is a notable shift of the BP maximum towards higher wavenumbers. As noted above, this is a common outcome of pressure application observed in previous works [6-9]. The pressure dependence of the position of the BP maximum is illustrated in Fig. 2 as semi-filled circles for pressure loading and as open circles for pressure release. The corresponding dependence of the two lowest energy optical vibrational modes [20] (rigid layer modes, RLM) of the crystalline counterpart is also presented in this figure for comparison. Apart from frequency and intensity changes, the application of pressure seems to have an unambiguous effect on BP's profile. In essence, the peak acquires a more symmetric shape with increasing pressure.

4. **Discussion**

In order to understand the pressure effects it would be helpful to briefly refer to some important structural information related to g-$As_2S_3$. The crystal of $As_2S_3$ is a layered-structure (2D) molecular crystal in the Raman spectrum of which both intralayer and interlayer vibrational modes are present. The former correspond to covalent bond stretching and bending modes of the constituent pyramidal units, while the latter (RLM) are manifestations of interlayer (van der Waals type) interactions. Sufficient experimental evidence supports that this type of structure is maintained, albeit loosely, even in the glass [19, 21]. In the frequency where the RLM are located, the Raman spectrum of the glass is dominated by the Boson peak. This was used in early studies as evidence of a common microscopic origin of the Boson peak and the RLM [22]; a similar suggestion was put forward for the isomorphous glass $As_2O_3$ [3]. Therefore, the comparison between the pressure dependence of the crystal and the glass is expected to reveal important information on the nature of the low energy excitations in this glass.

From the theoretical point of view pressure effects on the Boson peak have mainly been pursued in the framework of the soft potential model (SPM). Three different groups have developed models concerning the pressure dependence of the BP spectral features having as their basis soft excitations in amorphous media. (i) Klinger [13] has adopted the idea that two types of vibrational excitations contribute to the BP in a glass at ambient pressure: quasi-localized soft-mode vibrational excitations (atoms in soft atomic configurations), and acoustic modes. These two contributions are expected to have a different pressure dependence and hence different contribution to the BP intensity. Two characteristic pressure scales are involved; the soft mode one ($P_{sm}$ ~ 10 GPa) is associated with changes of the soft modes, while the other ($P_{ac}$ ~ 100 GPa) marks the limit for observing pressure effects in acoustic dynamics. The theory predicts that the BP frequency shifts to higher energies upon pressure application and its intensity accordingly drops. The rate of pressure increase of the BP is estimated as ~$10^{-5}$ THz bar$^{-1}$ (3.34 cm$^{-1}$ / GPa). As regards the acoustic modes the theory predicts an insensitivity of the BP frequency for pressure up to $P_{sm}$. (ii) Another theoretical approach built upon the SPM has been proposed by Hizhnyakov *et al.* [12], the basic idea of which is the conversion of soft localized modes to tunnelling two level systems under pressure application. An analytical expression that describes the frequency and pressure dependence of the BP lineshape was also provided and applied in the case of a polymeric glass. (iii) In the most powerful approach Gurevich, Parshin, and Schober (GPS) [10] attributed the appearance of the BP to a vibrational instability of weakly interacting quasi-localized harmonic modes. The frequency dependence of the VDoS of the *excess* modes predicted by the model entails a steep rise proportional to $\omega^4$ followed by a milder increase as $g(\omega) \propto \omega$ above a crossover frequency identified with the BP frequency. In the approximation of a Lorentzian distribution for both the random forces and the deformation potential the model resulted in a simple analytical expression concerning the pressure dependence of the BP frequency, which reads as:

$$\Omega_{BP}^{\max}(P) = \Omega_{BP}^{\max}(0)\left(1+\frac{P}{P_0}\right)^{\frac{1}{3}}. \tag{1}$$

$\Omega_{BP}^{\max}(P)$ is the frequency of BP maximum at pressure $P$, and $P_0$ is a parameter of the model which depends on the compressibility, the deformation potential and the magnitude of the total random forces acting on the quasi-local vibrations under pressure. The advantage of Eq. (1) is its straightforward comparison with experimental data. GPS [10] attempted a comparison of this equation with available experimental data and found a rather good agreement between theory and experiment.

However, they stressed that detailed experimental data at much higher pressures are needed in order to better appreciate the validity of the theory.

The systematic recording of Raman spectra of $As_2S_3$ shown in Fig. 1 makes it possible to test partly the predictions of Klinger's model [13] and in more detail the GPS model up to a pressure of 10 GPa. With the aid of local fits around the maximum of the BP the quantity $\Omega_{BP}^{max}(P)$ was estimated and is illustrated in Fig. 2. The rate of pressure increase of BP position is found to be ~10 cm$^{-1}$/GPa which is by a factor of three higher than the prediction of Ref. 13. The fit of Eq. 1 on the data of pressure loading is shown as a solid line revealing a very satisfactory agreement between experiment and theory over a wide range from ambient pressure up to 10 GPa; resulting in $\Omega_{BP}^{max}(0)$ = 27.5±1.1 cm$^{-1}$ and $P_0$ = 0.55±0.08 GPa. The BP position exhibits a strong pressure effect up to ~4 GPa followed by a moderate dependence at higher pressures. This is presumably associated to locally favored 2D structures interacting via weak van der Waals forces as will be discussed below in conjunction with the pressure dependence of the BP intensity.

The comparison between the BP frequencies for pressure loading and release (Fig. 2) reveals a hysteresis effect, which is typical in pressure studies. However, although the load-release pressure cycle seems to be not step-by-step (completely) reversible procedure, the $P = 0$ spectral features of the BP – before and after pressure application – seem to be indistinguishable within experimental error. It is interesting to notice that the tendency of the glass to keep memory of its high-pressure structure is appreciable in the pressure region below ca. 4 GPa.

Figure 1 reveals, apart from frequency, prominent intensity changes of the BP upon glass compression. Pressure-induced changes in BP intensity are qualitatively similar to those predicted by theories (see Fig. 1 in [10]). All experimental and theoretical/simulations works so far dealing with BP changes under pressure conclude a decrease of its intensity with increasing pressure. However, conclusions associated with excess of VDoS changes must be considered with care. BP originates from a VDoS spectrum in *excess* to the Debye contribution $g^D(\omega)$ and is therefore important (as has been stressed in Ref. [23]) not to overlook – as is often the case – the pressure dependence of $g^D(\omega)$ which is dictated by the pressure variation of the sound velocities. Since sound velocities increase with pressure (an exception to this is silica [24]), $g^D(\omega)$ decreases upon pressure application. As a result the pressure dependence of the excess $g^{exc}(\omega) = g(\omega) - g^D(\omega)$ cannot be evaluated since both $g(\omega)$ and $g^D(\omega)$ decrease with increasing pressure.

In addition, another parameter that is significant when intensity changes of the BP are considered is the frequency dependence of the first order (harmonic approximation) Raman spectrum that under certain approximations can be described by the following relation [25]:

$$I(\omega) \propto C(\omega) g(\omega) \frac{n(\omega,T)+1}{\omega} \qquad (2)$$

where $C(\omega)$ is the Raman coupling coefficient that is proportional to the scattering cross section of a vibrational mode at frequency $\omega$, and $n(\omega,T) = [\exp(\hbar\omega/k_B T) - 1]^{-1}$ is the Bose-Einstein occupation number. At sufficiently low frequencies the $[n(\omega)+1]/\omega$ factor is very well approximated by the term $1/\omega^2$, while in many glasses $C(\omega)$ is proportional to $\omega$ up to $\Omega_{BP}^{max}$ [26].

The excess spectrum, which is situated on the Debye VDoS, is expected to shift to higher energies due to the "strengthening" of the involved vibrational modes under pressure. Assuming that its intensity remains constant, the observed experimental intensity in the recorded spectra is modulated by the $C(\omega)/\omega^2$ factor and hence will appear reduced in so far as the excess peak shifts to higher $\omega$ under pressure application. Obviously, both $1/\omega^2$ and $C(\omega)/\omega^2 \propto \omega^{-1}$ point to an artificial reduction of the intensity of the excess modes. To quantify the above ideas we present in Fig. 3 the pressure dependence of the BP intensity. The dashed and solid lines represent the expected (artificial) suppression of the BP intensity (assuming a constant $g^{exc}(\omega)$ which shifts to higher frequencies) in the case where the prefactor of the VDoS in Eq. (2) is $1/\omega^2$ and $1/\omega$, respectively. The solid line also stands for the suppression of the BP intensity considering the $1/\omega$ dependence of excess VDoS predicted by the SPM [10]. Should the experimental data were located below these curves, we could support that the BP intensity, i.e. the excess VDoS, was really suppressed upon pressure application. Figure 3 reveals that the experimental data fit well the dashed line up to ~4 GPa. In conclusion, there is no evidence of considerable $g(\omega)$ suppression with pressure application. On the contrary there may be a case that the opposite effect occurs taking into account that $g^D(\omega)$ decreases with increasing pressure.

The same conclusion can be reached by comparing the pressure decrease of the quantities $g(\omega)$ and $g^D(\omega)$ for $As_2S_3$ glass. The former was estimated from the Raman spectra by measuring the intensity at constant energy, e.g. 30 cm$^{-1}$ (modulated by the coupling coefficient) and the latter from the pressure dependence of sound velocities estimated by Brillouin shifts [27] and the corresponding values of the refractive index as a function of pressure [17]. Figure 4 shows the pressure dependence of the normalized ratio $g(\omega)C(\omega)/g^D(\omega)$ at $\omega=30$ cm$^{-1}$. This ratio is a monotonically increasing function of pressure indicating an enhancement of the $g(\omega)C(\omega)$ quantity with respect to the Debye level. Since $C(\omega)$ is expected to decrease as pressure increases, we conclude that part of the $g(\omega)C(\omega)$ suppression comes from the second term. The inset shows that the excess, i.e. the difference between the two aforementioned quantities increases with increasing pressure questioning the general perception of a strong pressure-induced suppression of the BP.

As has been mentioned above the characteristic pressure ~4 GPa is associated with four experimental observations: (i) a change of the rate of the BP frequency increase (Fig. 2), (ii) the onset of the hysteresis effects with pressure release (Fig. 2), (iii) a sudden increase of the intensity of the BP (Fig. 3), and (iv) a sudden change of the $g(\omega)C(\omega)/g^D(\omega)$ ratio (Fig. 4); which imply the onset of some structural transition. In the crystal, a dimensionality transition 2D→3D commences in the pressure region 3-4 GPa [20] as is also exhibited by the frequency ratio of the RLM (see inset in Fig. 3). A similar effect on the structure of glass may be expected leading essentially to a *glass-to-glass transition* under pressure in view of the partial preservation of the layered structure in the glass. This kind of transition would be a reasonable rationale for the two aforementioned changes of the frequency and intensity of the BP around 4 GPa.

Closing this section we would like to comment on some remarkable changes of the BP lineshape under pressure. Comparing for example the BP lineshapes at P=0 and P=3.6 GPa, Fig. 5 (left panel) we discern a drastic change in its asymmetry, i.e. the BP peak becomes much more symmetric at high pressures. Similar effects have been recently observed [23] in analyses of the spectral details of the excess VDoS. Indeed, a change of the $g^{exc}(\omega)$ shape from a frequency distribution with convex-like tails to a distribution with concave-like tails can indeed induce such drastic changes in the frequency-reduced VDoS $g(\omega)/\omega^2$ which is proportional to the experimental Raman spectrum. In Fig. 5 (right panel) we demonstrate this by choosing two types for the excess VDoS with concave and convex tails superimposed on a Debye contribution. On quantitative grounds, the change of the excess distribution tails from convex to concave is capable of reducing the asymmetry of the BP and shifting the frequency of its maximum to much higher energies. This is an alternative way of accounting for the observed changes in high pressure Raman studies. However, more detailed analysis of experimental data and maybe theoretical wok on the true shape of the excess VDoS and its dependence upon pressure is needed to clarify the above issues.

## 5. Conclusions

A detailed Raman spectroscopic study on glassy $As_2S_3$ has been conducted as a function of pressure. Particular emphasis was placed on the pressure effects of Boson peak's spectral details. The main conclusions of the present study can be summarized as follows:

(i) Pressure induces a noticeable shift of $\Omega_{BP}^{max}$ towards higher energies. The pressure dependence of $\Omega_{BP}^{max}$ can be very well described by the predictions of the soft potential model recently formulated by Gurevich *et al*. [10(b)].

(ii) Pressure-induced changes in the Boson peak intensity are more complicated. According to a naive interpretation which is usually adopted in such studies pressure suppresses the BP intensity. However, as we have pointed out in this paper, one should not overlook the corresponding pressure dependence of the Debye VDoS. Taking this into account we found that experimental data not only do not exhibit a BP intensity suppression but on the contrary the Debye contribution was found to decrease faster than the total VDoS, thus implying a net increase of the excess VDoS from which the BP originates.

(iii) A more symmetric BP shape is observed at high pressures. This observation can be possibly associated to the very details of the frequency distribution of the excess VDoS. Slight modifications of the latter may result on significant alterations of the associated artificially produced asymmetric peak in the frequency-reduced representation, namely, the BP.

Finally, substantial experimental evidence indicates a glass-to-glass transition of $As_2S_3$ that occurs at ~ 4GPa. This transition is most likely related to structural alterations of the system and in particular to a change of the local network dimensionality form 2D to 3D as it also occurs in the corresponding crystal.

**Acknowledgements**

K.S.A. and S.N.Y. acknowledge illuminating discussions with Dr. H. R. Schober. K.S.A. acknowledges financial support from the project "PYTHAGORAS I" funded by the Greek Ministry of National Education and Religion Affairs.

**Figure Captions**

**Fig 1:** Raman spectra of $As_2S_3$ as a function of pressure. Integrated intensity corresponding to the [270-420] $cm^{-1}$ spectral region was used in order to normalize the spectra obtained at different pressures.

**Fig. 2:** The BP maximum as a function of pressure (filled circles for the procedure of pressure application and half-filled ones for the pressure release). Solid line indicates the theoretical prediction after fitting the experimental data using Eq. (1). Diamonds denote the corresponding dependences of the orpiment crystal RLM . Dotted lines refer to the pressure dependence of the two RLM of the crystal and are used as guides to the eye. Inset shows the calculated ratio of the two RLM frequencies as a function of pressure.

**Figure 3.** Plot of the position of normalised BP maximum at different pressures indicating its apparent intensity suppression at least up to the pressure of ~4GPa. Solid and dashed lines show the normalised $1/\omega$ and $1/\omega^2$ curves respectively.

**Figure 4.** Plot of the ratio of $g(\omega)C(\omega)/g^D(\omega)$ as a function of pressure indicating an enhancement of the $g(\omega)C(\omega)$ quantity with respect to the Debye level. Inset shows plots of the $g(\omega)C(\omega)$ normalized values (resulting from the recorded spectra at 30$cm^{-1}$) and the corresponding normalized values of $g^D(\omega)$ (obtained from Brillouin data and refractive index measurements).

**Figure 5.** Spectra of $As_2S_3$ at ambient conditions (a) and at 3.6GPa (b), magnified so that they cover the spectral range of BP. Shadowed areas in (c) and (d) represent demonstrations of symmetric concave-tailed and convex-tailed excess of VDoS, $g^{exc}(\omega)$, respectively. We can approximate the estimated experimental Raman spectrum produced by these excess VDoS if we sum $g^{exc}(\omega)$ to $g^D(\omega)$ and then calculate the quantity $g(\omega)/\omega^2$. For the case of a concave-tailed $g^{exc}(\omega)$ a highly asymmetric BP appears in the estimated Raman spectrum that resembles the BP seen in (a). In contrast, the convex-tailed $g^{exc}(\omega)$ results in a more symmetric BP which corresponds to the high pressure BP of $As_2S_3$ shown in (b).

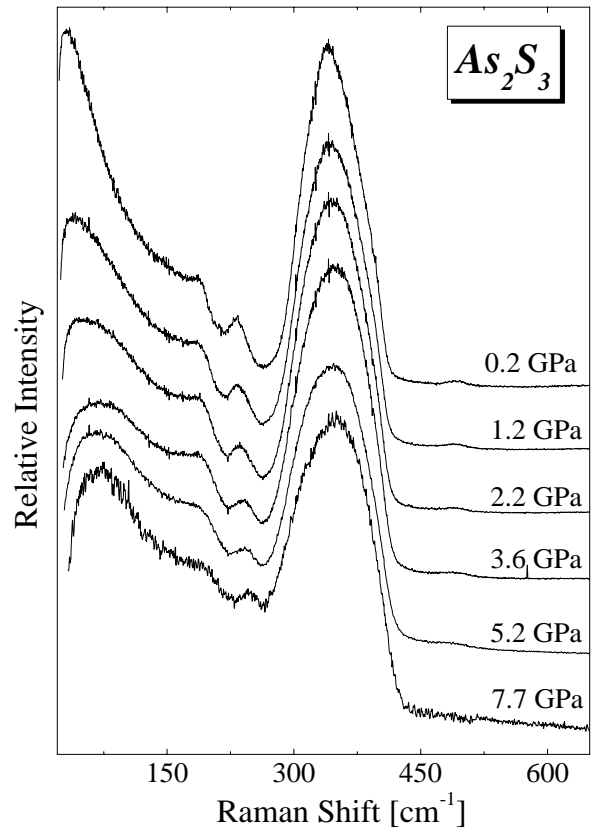

**Fig. 1**

*Pressure Dependence of the Boson...*

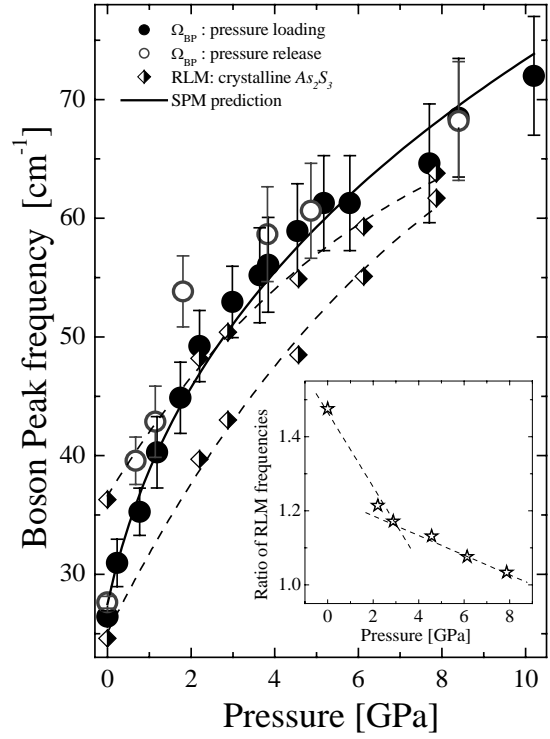

**Fig. 2**

*Pressure Dependence of the Boson…*

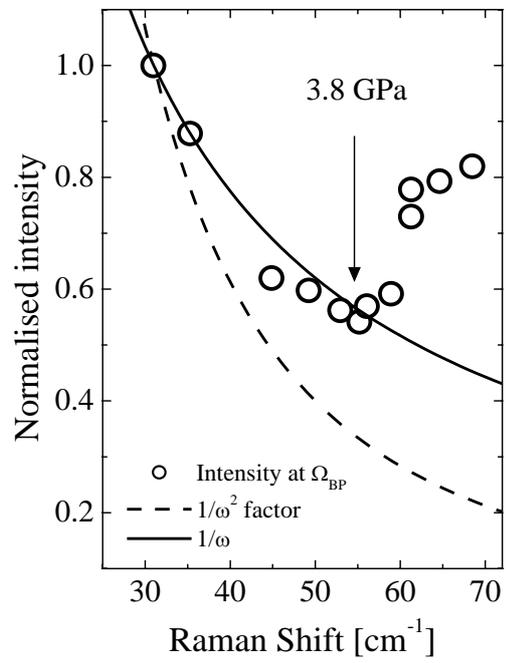

**Fig. 3**

*Pressure Dependence of the Boson…*

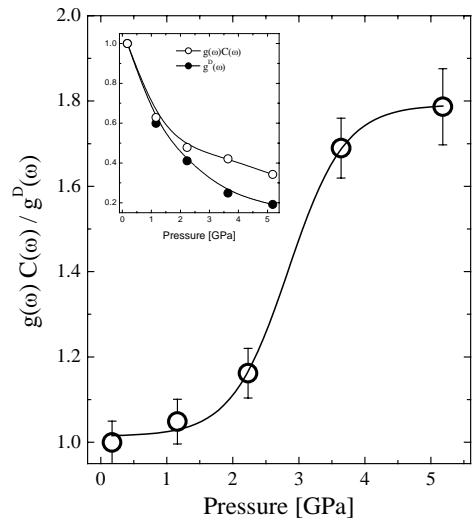

**Fig. 4**

*Pressure Dependence of the Boson…*

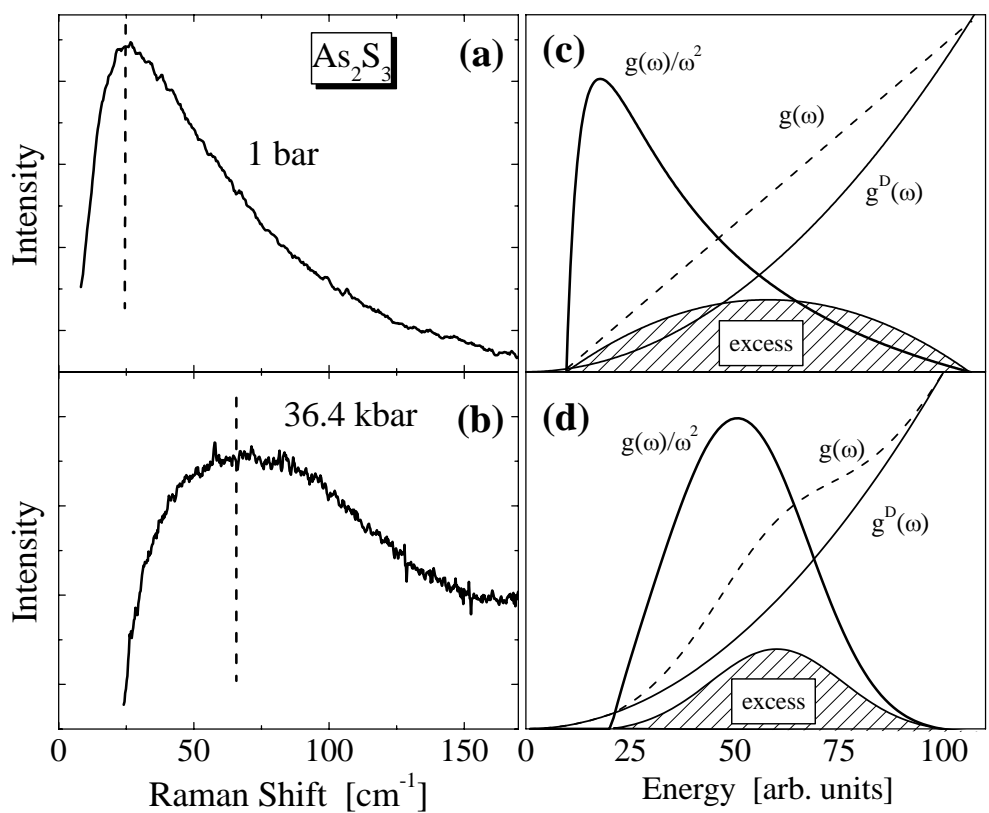

**Fig. 5**

*Pressure Dependence of the Boson...*